\documentclass[twoside]{ilcws10}
\usepackage[latin1]{inputenc}
\usepackage[dvips]{graphicx,epsfig,color}
\usepackage{wrapfig,rotating}
\usepackage{amssymb,amsmath,array}

\begin{document}
\title{
%%%%   Paper title goes here  %%%%%%%%%%%%%%
ISIS2: Pixel Sensor with Local Charge Storage for ILC Vertex Detector} %%
%***********************************************************************
% AUTHORS INFORMATION AREA
%***********************************************************************
\author{Yiming Li$^1$, Chris Damerell$^2$, Rui Gao$^1$, Rhorry Gauld$^1$, Jaya John John$^1$,\\
Peter Murray$^2$, Andrei Nomerotski$^1$, Konstantin Stefanov$^3$, Steve Thomas$^2$, \\
Helena Wilding$^1$, Zhige Zhang$^2$
% Optional short acknowledgment: remove next line if non-needed
%\thanks{Do we need to add the funding source acknowledgment here???}
% DO NOT MODIFY THE FOLLOWING '\vspace' ARGUMENT
\vspace{.3cm}\\
% Addresses and institutions (remove "1- " in case of a single institution)
1- Sub-department of Particle Physics \\
University of Oxford, Oxford OX1 3RH - UK
% Remove the next three lines in case of a single institution
\vspace{.1cm}\\
2- Rutherford Appleton Laboratory - UK\\
3- Sentec Ltd - UK
}
%%***********************************************************************
% END OF AUTHORS INFORMATION AREA
%***********************************************************************

\maketitle

\begin{abstract}
ISIS (In-situ Storage Imaging Sensor) is a novel CMOS sensor with multiple charge storage capability developed for the ILC vertex detector by the Linear Collider Flavour Identification (LCFI) collaboration. This paper reports test results for ISIS2, the second generation of ISIS sensors implemented in a 0.18 micron CMOS process. The local charge storage and charge transfer were unambiguously demonstrated.
\end{abstract}

\section{Introduction}
\begin{wrapfigure}{r}{0.5\columnwidth}
\centerline{\includegraphics[width=0.45\columnwidth]{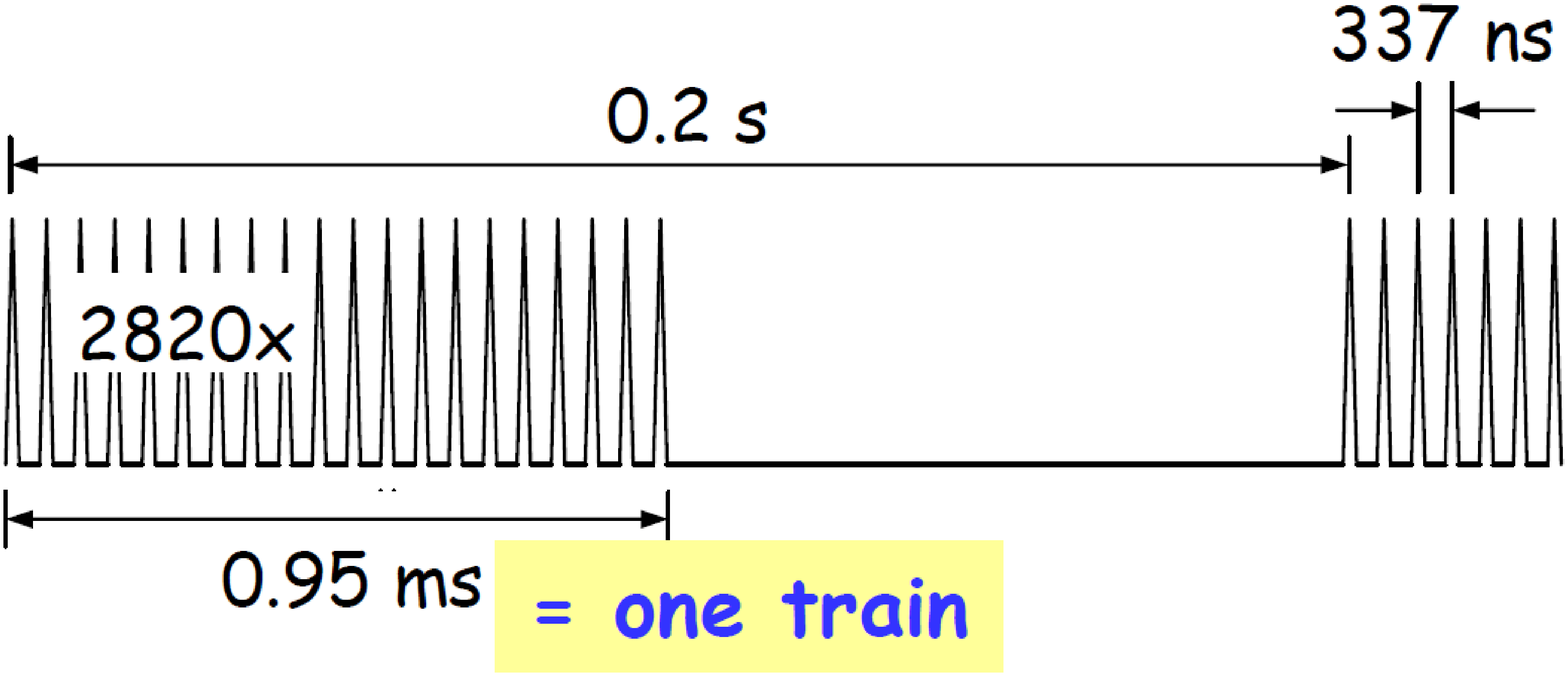}}
\caption{ILC bunch train structure.}\label{Fig:ILCtimeslicing}
\end{wrapfigure}

%ILC vertexing requirements
In order to achieve the precision targeted at the ILC, the vertex detector must meet challenging requirements. The resolution of $\sim 3 \mu m $ is a straightforward one to enable the high precision studies of processes requiring b-tagging such as the measurement of top quack forward backward assymmetry, Higgs self-coupling etc~\cite{SiDLOI}. In the $e^+e^-$ machine huge amount ($10^5\sim10^6$ per collision) of $e^+e^-$ pairs will be present as background due to beamstrahlung. Figure \ref{Fig:ILCtimeslicing} shows the ILC beam structure. It is essential for the vertex detector to keep the occupancy below $1\%$. Therefore time slicing is necessary. In addition, the material budget will be $0.1 X_0\%$ per layer, which largely limits the cooling option to air cooling.

Two different approaches exist to tackle the low occupancy issue. The first is fast-readout during the collisions, at least 20 frames in $\sim 1 ms$ time. This has been pursued by Column Parallel CCDs~\cite{CPCbib1, CPCbib2, CPCbib3}, MAPS~\cite{MAPSbib1,MAPSbib2} and DEPFET~\cite{DEPFETbib1,DEPFETbib2} sensor types just to name a few. The ISIS sensor pursues another approach, which is to store the charge at the pixel level during the bunch train and to read out during the quiet time. This reduces the required peak power and avoids power cycling, which is needed for the first approach.

\subsection{ISIS principle}

\begin{figure}[htbp]
\centerline{\includegraphics[width=0.9\textwidth]{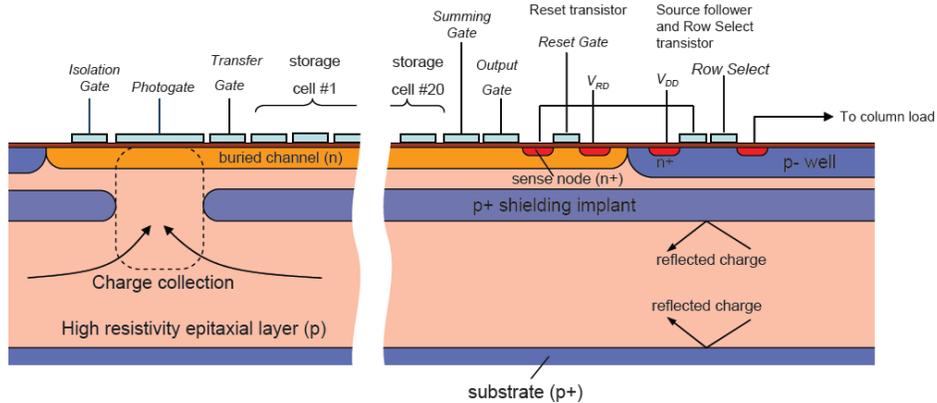}}
\caption{ISIS design concept.}\label{Fig:BC-reset-transistor}
\end{figure}

The design concept of ISIS sensor can be illustrated in the cross section of a pixel as in Figure \ref{Fig:BC-reset-transistor}. During the collisions the charge generated in the $\sim 20 \mu m$ epitaxial layer is collected under the photogate. Then the raw charge is transferred down to the pixel-level storage cells, equivalent to a short CCD column. The signal charge will be shifted to the register every $50 \mu s$ as the time slicing required. The charge-voltage conversion and readout of full array will happen in the $~200 ms$ quiet time between the bunch trains. The CCD register is protected from the epitaxial layer by the deep $p^+$ implant.

\subsection{History}
Fast framing CCD optical cameras based on ISIS principle has been under development since around ten years ago~\cite{GojiEtoh}, with the maximum frame rate of $\sim 100 Megaframes/s$ achieved. The ISIS sensor as a particle detector for ILC vertexing was pioneered by Linear Collider Flavour Identification (LCFI) collaboration in 2003.

\begin{wrapfigure}{r}{0.4\columnwidth}
\centerline{\includegraphics[width=0.35\columnwidth]{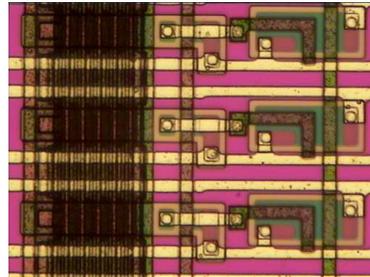}}
\caption{Three pixels on ISIS1}\label{Fig:ISIS1}
\end{wrapfigure}

The first generation device ISIS1 was produced to prove the feasibility of local charge storage on a relatively large CCD pitch. A example of ISIS1 pixels is shown in Figure \ref{Fig:ISIS1}. The ISIS1 sensors was manufactured by e2v technologies~\cite{e2v} with $\sim 2 \mu m $ CCD process. There are 5 storage cells for each pixel, each with area of $160 \times 40 \mu m^2$. They were successfully tested with X-ray and testbeams~\cite{ISIS1bib1,ISIS1bib2,ISIS1bib3}.

ISIS2, the second generation of ISIS sensor was designed and manufactured on a smaller pitch and with more storage cells. The ISIS2 sensors were received in 2008 and the main results of their testing will be summarized in the following sections. Section \ref{sec:design} will explain the design of ISIS2 in detail. Section \ref{sec:test-result} will report the testing results on test structure and main arrays.

\section{ISIS2 Design}\label{sec:design}

The ISIS2 sensors were manufactured in a $\sim 0.18 \mu m$ CMOS process by Jazz Semiconductor~\cite{Jazz}. The small feature size of the foundry enables 20 storage registers every pixel, each with the area of $3 \times 5 \mu m^2$  (comparing to the $20 \times 40 \mu m^2$ storage cell on ISIS1).

\begin{figure}[htbp]
\centerline{\includegraphics[width=0.9\textwidth]{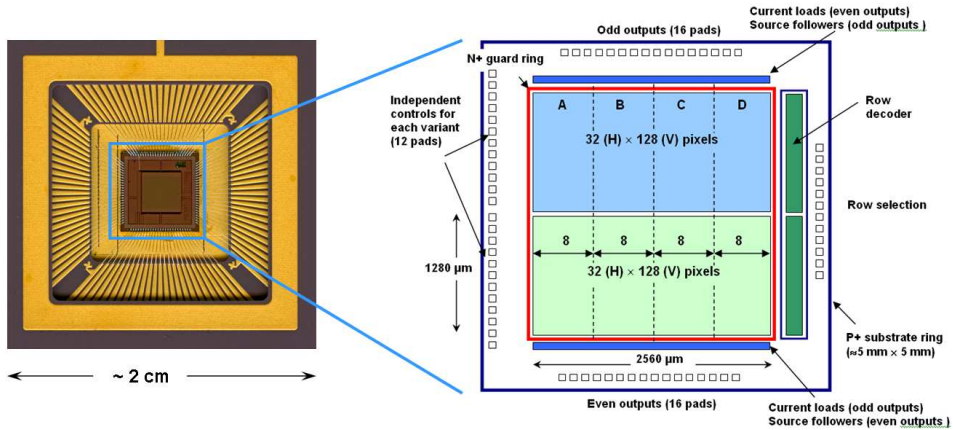}}
\caption{ISIS2 sensor wired-bonded in a ceramic package(left) and ISIS2 floor plan(right).} \label{Fig:package-floorplan}
\end{figure}

Figure \ref{Fig:package-floorplan} shows a photo of a packaged ISIS2 sensor and its floor plan. The total size of the sensor is $5 \times 5 mm^2$. The sensor has 32 columns and 128 rows, divided into the upper and lower halves, with buried-channel and surface-channel reset transistor respectively. The columns are equally divided into four sections featured by different deep $p^+$ well variations. The readout from 8 columns in each sections are multiplexed, and the rows are controlled by a rolling shutter.

\begin{figure}
\centerline{\includegraphics[width=0.6\textwidth]{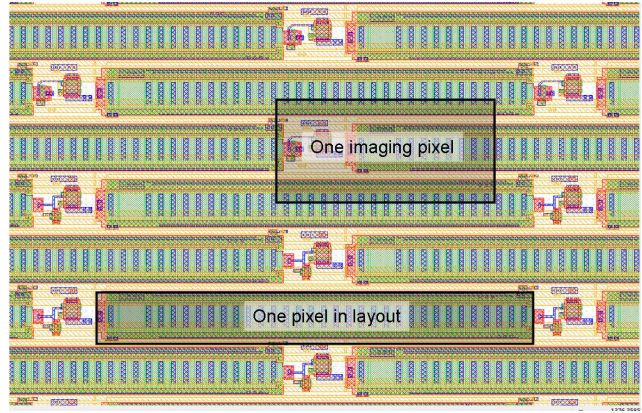}}
\caption{ISIS2 pixel layout. An $80 \times 10 \mu m^2$ pixel and an imaging pixel of the size $40 \times 20 \mu m^2$ are in shadow.}\label{Fig:pixel-layout}
\end{figure}

\begin{figure}[htbp]
\centerline{\includegraphics[width=0.9\textwidth]{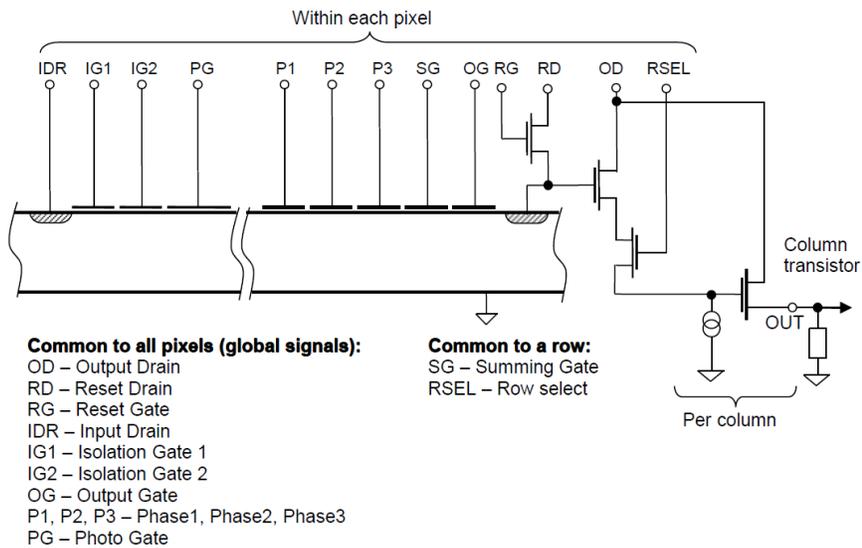}}
\caption{Diagram of an ISIS2 pixel.}
\label{Fig:pixel-diagram}
\end{figure}

The ISIS2 imaging pixel size is $40 \times 20 \mu m^2$ while the pixel layout size is $80 \times 10 \mu m^2$ as in Figure \ref{Fig:pixel-layout}. This miniaturization is a major progress compared to ISIS1 thanks to the small feature size provided by the foundry. Each pixel of ISIS2 contains a 3-phase CCD with 20 storage cells, a reset transistor, source follower and a row select transistor. There is a charge injection input next to the photogate and this controlled charge injection is useful for testing. All clock and bias signals are shared between all pixels except the Summing Gate (SG) and Row Select (RSEL) signals, which can be controlled by a separate transfer gate at row level. A diagram of the pixel structure described here is shown in Figure \ref{Fig:pixel-diagram}.

\begin{wrapfigure}{r}{0.6\columnwidth}
\centerline{\includegraphics[width=0.57\columnwidth]{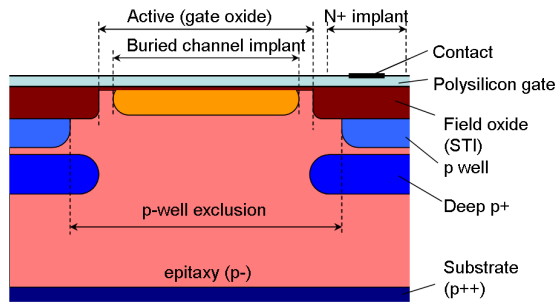}}
\caption{ISIS2 pixel cross-section under the photogate.}
\label{Fig:pwell-xsec}
\end{wrapfigure}

The separation of the buried channel and the epitaxial layer with the deep $p^+$ implant is another improvement after ISIS1 only possible in a modern CMOS process. Figure \ref{Fig:pwell-xsec} shows the part of the cross-section of an ISIS2 pixel under the photogate. The opening of the deep $p^+$ implant is shown (not to scale) through which the deposited charge is collected. In order to investigate the deep $p^+$ implant, four sections of the ISIS2 chip include following variations: (1)without deep $p^+$; (2)with deep $p^+$ but without the opening under the photogate; (3) with deep $p^+$ and the opening; (4) with deep $p^+$ and a larger opening compared to (3).

The fabrication process enables a range of additional ISIS2 variations which can be helpful in studying the performance optimization. These variations are listed below:
\begin{itemize}
\item{Deep $p^+$ well (as mentioned above);}
\item{Reset transistors implemented in surface channel and buried channel;}
\item{Different CCD gate and gap width;}
\item{Change in dopant concentration up to 20\%}
\end{itemize}

\section{ISIS2 Test Results}\label{sec:test-result}

\subsection{Test Structure}
\begin{figure}[htbp]
\centerline{\includegraphics[width=0.6\textwidth]{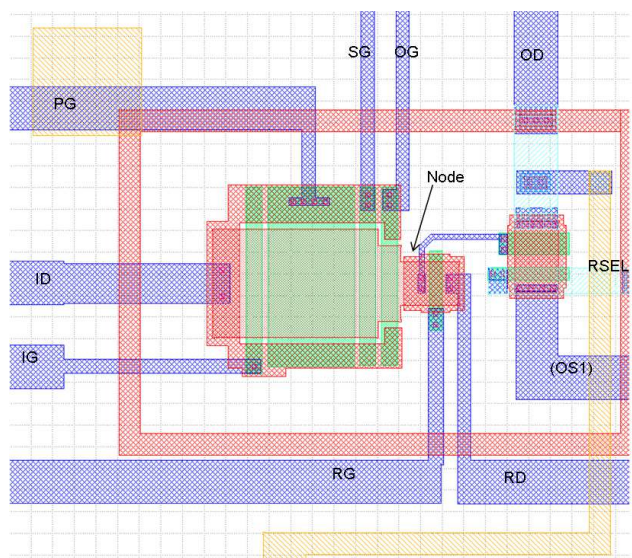}}
\caption{ISIS2 test structure.}
\label{Fig:test-structure}
\end{figure}

A simple test structure (as shown in Figure \ref{Fig:test-structure}) was included on the same ISIS wafer. It is similar to a pixel in the main array except for the absence of the storage CCD cells. This structure allows us to establish the optimum operation point before studying the full array. Over half of the test structure area is taken by the $4 \times 5 \mu m^2$ photogate. The small size of the output node will result in its small capacitance, hence a large charge to voltage conversion factor and very low noise. On the other hand the small size together with the tapered shape from the output gate to the output node could complicate charge transfer between the two and also means that the edge effects and 3D fringing fields can become important.

\begin{figure}[htbp]
\centerline{\includegraphics[width=0.6\textwidth]{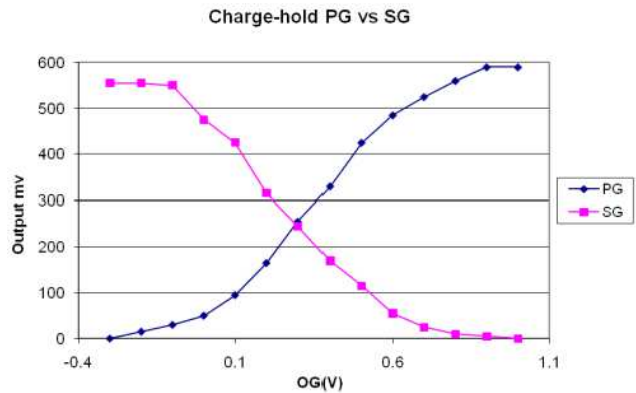}}
\caption{Charge that can be held under summing gate(SG) and that leaks to output node(PG) at different OG voltage. \cite{Vertex09ISIS}}
\label{Fig:fringe-effect}
\end{figure}

%~~~~~~ Fringing field ~~~~~~
It turned out that the fringing effect is important because the voltage on the output gate (OG) is pulled up by the constant 5 V voltage on the neighbouring output gate. It was found that the OG bias should set to below 0 V otherwise the charge will leak from the summing gate directly into the output node. The OG bias value for a typical CCD could be above 1 V. Figure \ref{Fig:fringe-effect} shows an example of such behaviour \cite{Vertex09ISIS}. The sensor was illuminated with LED light and then the photogate was lowered while the summing gate was kept high. The charge collected should be temporarily held within summing gate if the OG functions properly. The line labeled "SG" shows the charge kept under summing gate and "PG" shows the charge that leaked through OG to the output node. It is clear that for a low OG voltage the charge can be transferred as expected while at high OG the transfer is severely hampered by fringe field and happens prematurely.

%~~~~~~ Slow readout ~~~~~
\begin{wrapfigure}{r}{0.5\columnwidth}
\centerline{\includegraphics[width=0.45\columnwidth]{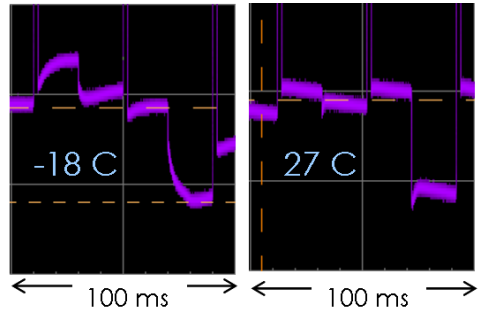}}
\caption{Signal output at $-18^{\circ}C$ and $27^{\circ}C$.}
\label{Fig:slow-readout}
\end{wrapfigure}
Another unexpected feature discovered in ISIS2 is the highly resistive polysilicon gate. The gates were left unintentionally undoped which caused the high resistivity. The immediate result is that the time constant to charge up the gate to a certain voltage is substantially longer than expected, and a charge transfer between two neighbouring gates typically takes $\sim ms$. The time constant also varies with the temperature. From Figure \ref{Fig:slow-readout} which shows the voltage output it is clear that the response at lower temperature is much slower. Though the time of operation was largely limited by the charge transfer speed, it was also found that the charge survives up to a few seconds in the sensor which enables us to investigate charge transfers in the CCD structure and study the transfer efficiency.

\begin{figure}[htbp]
\centerline{\includegraphics[width=0.8\textwidth]{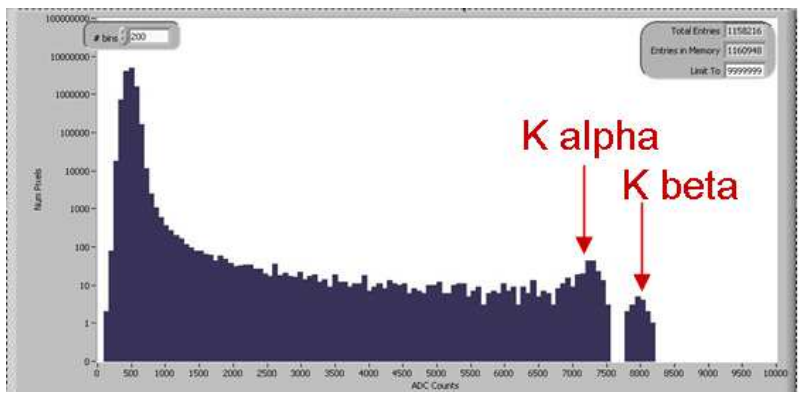}}
\caption{Amplitude spectrum of $^{55}Fe$ source. The left peak is the pedestal and the two labelled signal peaks on the right are caused by the $^{55}Fe$ X-rays. These are the direct hits on output node at $31^{\circ}C$.}
\label{Fig:Fe55-calibration}
\end{figure}

%~~~~~~~ Fe55 calibration ~~~~~~~~~
The absolute calibration of the test structure is done using a standard $^{55}Fe$ radiative source. A minimum noise of $6 e^-$ allows the $K_{\alpha}$ and $K_{\beta}$ peaks to be resolved, which correspond to $1620 e^-$ and $1780 e^-$ generated in the epitaxial layer, as shown in Figure \ref{Fig:Fe55-calibration}. It can be derived that the response of the output node is $24 \mu V/e^-$ as designed.

With the X-ray source it is possible to estimate the charge transfer efficiency (CTE) on the test-structure. The output from two cases are compared: (1) direct X-ray hits on the output node and (2) hits collected on the photogate and transferred through summing gate to the output node. Separation of the $K_{\alpha}$ peak and pedestal peak in both cases give the charge on the output node. The ratio of the charge in case (2) over case (1) is the CTE estimated as in Table \ref{Tab:CTE-test-structure}. This measurement was repeated at room temperature and low temperature, and the CTE estimation agrees well. This relatively low CTE is probably caused by the tapered shape of transition from the summing gate to the output node.
\begin{table}
\centering
\begin{tabular}{|c|c|c|}
\hline
Temperature & $-10^{\circ}C$ & $31^{\circ}C$ \\
\hline
CTE & 94.2\% & 94.5\% \\
\hline
\end{tabular}
\caption{CTE on test-structure.}
\label{Tab:CTE-test-structure}
\end{table}

\begin{figure}[htbp]
\centerline{\includegraphics[width=0.6\textwidth]{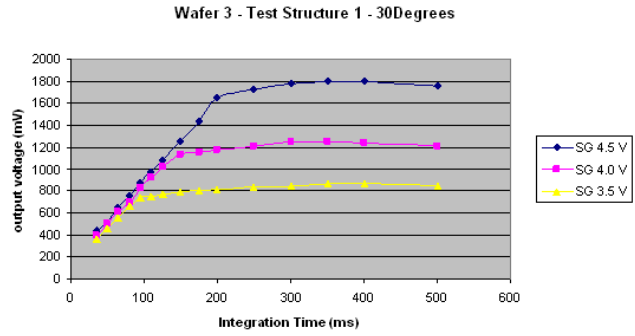}}
\caption{Dependence of the output voltage on the photogate integration time.}
\label{Fig:test-structure-linearity}
\end{figure}

%~~~~~~~ Charge transfer ~~~~~~~~~
In addition to the X-ray source, the charge transfer was also studied with charge injection, dark current integration and LED pulses. One example using dark current is given in Figure \ref{Fig:test-structure-linearity} which shows the dependence of the output voltage on the photogate integration time. From this measurement the summing gate capacity can be estimated to $5000 \sim 10000 e^-$, depending on the summing gate bias.

\subsection{Main Array}
\begin{figure}[htbp]
\centerline{\includegraphics[width=0.7\textwidth]{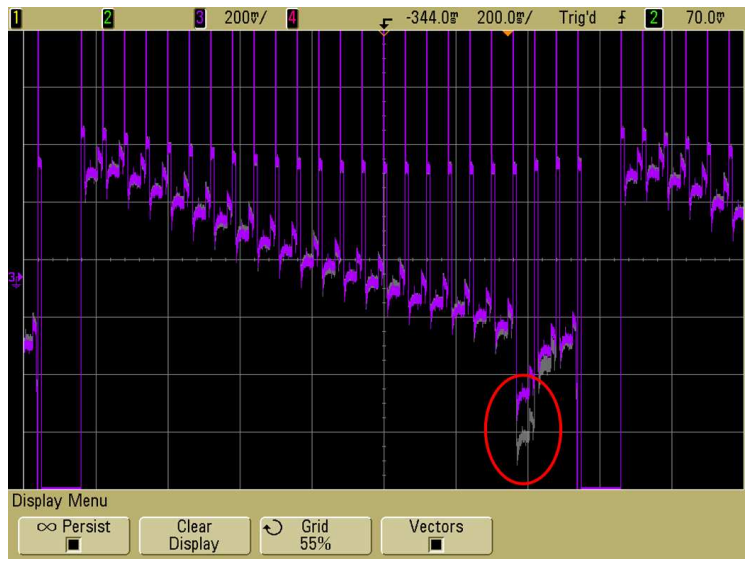}}
\caption{The oscilloscope trace of a successful charge transfer in the ISIS2 main array.}
\label{Fig:first-ISIS2-main-array}
\end{figure}

%~~~~~~~ Scope trace ~~~~~~~~~
An example of successful charge transfer in the ISIS2 main array is shown in Figure \ref{Fig:first-ISIS2-main-array}. The oscilloscope traces both with and without charge injection are shown. The difference emphasized with the circle is due to the output of the injected charge after being transferred down the 20 storage cells. The linear accumulation of leakage current on the twenty CCD registers is also clearly visible.

%~~~~~~~ Effort to minimize the readout time ~~~~
The slow readout speed mentioned earlier means that it takes a long time (up to seconds) to read a full pixel, during which time the dark current will cause a considerable shot noise and more worryingly, could saturate the well capacity. A lot of efforts therefore have been directed to minimize the readout time. The most straightforward way of reducing the dark current will be to reduce temperature, but this will lead to higher resistivity of the polysilicon gate, hence even longer readout time. The operation temperature between $-5^{\circ}C$ and room temperature was found to be optimal. After optimization the transfer time for the CCD gates was set to 85\% of the summing gate time. At room temperature a readout rate of about 10 Hz was achieved.

\begin{figure}[htbp]
\centerline{\includegraphics[width=0.8\textwidth]{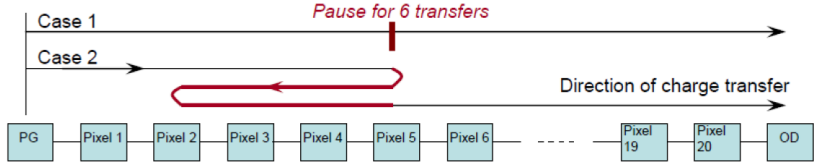}}
\caption{CTE measurement method using reversing of charge in the CCD register.}
\label{Fig:charge-reversal}
\end{figure}

%~~~~~~~ CTE measurement (1) ~~~~~~~~
In order to measure the CTE for the 20 storage cells with $^{55}Fe$ X-ray source, a challenge is to disentangle the hits on the photogate and those on the storage cells. There are two different approaches on this issue. Since three-phase CCD registers are used, the charge can be transferred in both directions. The first approach uses charge manipulation and compares the two cases as illustrated in Figure \ref{Fig:charge-reversal}. In case (2) the charge from photogate is first moved forward, then backward for $n$ cells, then forward again. In case (1) charge always moves forward, but will be paused in the middle simply to keep both sequences at the same length. Therefore there will be $2n$ more transfers in case (2) than in (1). Assume the outputs measured are $S_1$ and $S_2$, then $S_2/S_1 = CTE^{2n} \simeq 1 - 2n\times CTI $, where $CTI = 1 - CTE$ is Charge Transfer Inefficiency. This measurement gives $CTE \gtrsim 99\%$. The uncertainty of this measurement is quite large, mainly due to the temperature fluctuation during the measurements.

\begin{figure}[htbp]
\centerline{\includegraphics[width=0.8\textwidth]{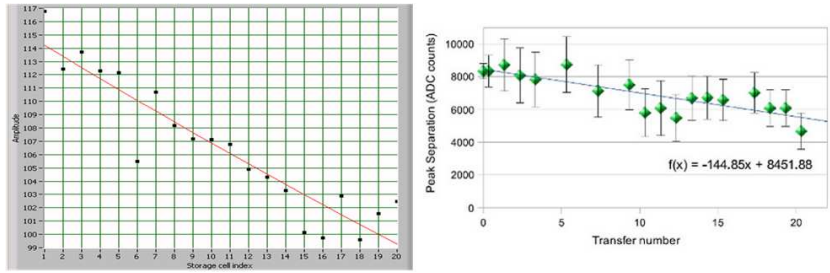}}
\caption{Position of $^{55}Fe$ signal peak versus number of transfers. The left plot was obtained with a moving $^{55}Fe$ source \cite{Vertex09ISIS} and the right with a mechanical shutter between the source and the sensor.}
\label{Fig:CTE-slopes}
\end{figure}

%~~~~~~~ CTE measurement (2) ~~~~~~~~
Another independent approach to measure the CTE is to define the time window for X-ray illumination. Two measurements based on this idea were made at different setups. One of them uses a moving X-ray source and the other use a fast mechanical shutter between the source and the sensor to control the timing of the X-ray illumination. If the initial charge package is $S_0$, after $N$ transfers the charge left will be $S_N = S_0 \times ( 1- CTI)^N \sim S_0(1- N\times CTI)$. So the CTE is calculated from the measured output signal dependence on the number of transfers as in Figure \ref{Fig:CTE-slopes}. The measurement with a moving source gives a CTE of 99.3\% \cite{Vertex09ISIS} and that with a shutter measures 98.4\%. It should be noted that these are measured under different conditions (temperature, frequency) and using different ISIS2 variations, so they are not directly comparable. Still these numbers are consistent within uncertainties and show a reasonably high transfer efficiency.

\begin{figure}
\centerline{\includegraphics[width=0.55\columnwidth]{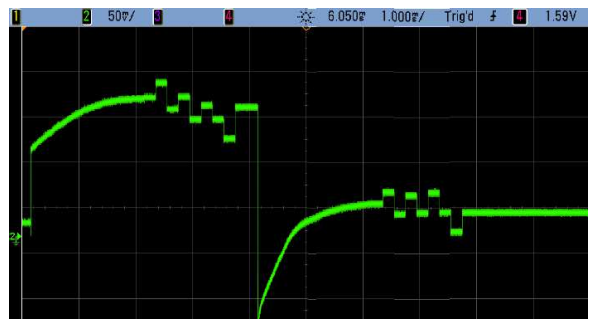}}
\caption{The voltage output of 8 columns multiplexed.}\label{Fig:multiplexed-output}
\end{figure}

%~~~~~~~ Readout ~~~~~~~~
As mentioned in Section \ref{sec:design}, all rows are read out by a rolling shutter and each eight columns are multiplexed. Figure \ref{Fig:multiplexed-output} shows an example of the multiplexed voltage output.

\section{Conclusion}
The tests of ISIS2 sensors showed successful charge storage and transfer abilities for a short CCD register implemented in a CMOS process. A few unexpected features such as the slow charge transfer caused by the resistive polysilicon gates were well understood and should be easy to
correct in future iterations. A large area ISIS sensor with more compact and optimized pixel geometry, and readout architecture with data serialization is a viable technology for the ILC vertexing.

% ****************************************************************************
% BIBLIOGRAPHY AREA
% ****************************************************************************

\begin{footnotesize}
% IF YOU DO NOT USE BIBTEX, USE THE FOLLOWING SAMPLE SCHEME FOR THE REFERENCES
% ----------------------------------------------------------------------------

% ----------------------------------------------------------------------------

\end{footnotesize}

% ****************************************************************************
% END OF BIBLIOGRAPHY AREA
% ****************************************************************************

\end{document}